\documentclass[twocolumn, twocolappendix]{aastex63}

\newcommand{\sgr}{SGR\,1935+2154}
\submitjournal{ApJL}
\shorttitle{Palomar Gattini-IR limits on NIR flares from \sgr}
\shortauthors{K. De et al.}
\graphicspath{{./}{figures/}}

\begin{document}

\title{Constraining the X-ray - Infrared spectral index of second-timescale flares from \sgr\ with Palomar Gattini-IR}

\correspondingauthor{Kishalay De}
\email{kde@astro.caltech.edu}

\author[0000-0002-8989-0542]{Kishalay De}
\affil{Cahill Center for Astrophysics, California Institute of Technology, 1200 E. California Blvd. Pasadena, CA 91125, USA}

\author{Michael C. B. Ashley}
\affil{School of Physics, University of New South Wales, Sydney NSW 2052, Australia}

\author[0000-0002-8977-1498]{Igor Andreoni} 
\affil{Cahill Center for Astrophysics, California Institute of Technology, 1200 E. California Blvd. Pasadena, CA 91125, USA}

\author[0000-0002-5619-4938]{Mansi M. Kasliwal}
\affil{Cahill Center for Astrophysics, California Institute of Technology, 1200 E. California Blvd. Pasadena, CA 91125, USA}

\author{Roberto Soria}
\affil{National Astronomical Observatories, Chinese Academy of Sciences, Beijing 100012, China}
\affil{Sydney Institute for Astronomy, The University of Sydney, Sydney, NSW 2006, Australia}

\author{Gokul P. Srinivasaragavan} 
\affil{Cahill Center for Astrophysics, California Institute of Technology, 1200 E. California Blvd. Pasadena, CA 91125, USA}

\author{Ce Cai}
\affil{Key Laboratory of Particle Astrophysics, Institute of High Energy Physics, Chinese Academy of Sciences, 19B Yuquan Road, Beijing 100049, China}
\affil{University of Chinese Academy of Sciences, Chinese Academy of Sciences, Beijing 100049, China}

\author{Alexander Delacroix}
\affil{Caltech Optical Observatories, California Institute of Technology, Pasadena, CA 91125, USA}

\author{Tim Greffe}
\affil{Caltech Optical Observatories, California Institute of Technology, Pasadena, CA 91125, USA}

\author{David Hale}
\affil{Caltech Optical Observatories, California Institute of Technology, Pasadena, CA 91125, USA}

\author[0000-0001-9315-8437]{Matthew J. Hankins}
\affil{Cahill Center for Astrophysics, California Institute of Technology, 1200 E. California Blvd. Pasadena, CA 91125, USA}

\author{Chengkui Li}
\affil{Key Laboratory of Particle Astrophysics, Institute of High Energy Physics, Chinese Academy of Sciences, 19B Yuquan Road, Beijing 100049, China}

\author{Daniel McKenna}
\affil{Caltech Optical Observatories, California Institute of Technology, Pasadena, CA 91125, USA}

\author{Anna M. Moore}
\affil{Research School of Astronomy and Astrophysics, Australian National University, Canberra, ACT 2611, Australia}

\author{Eran O. Ofek}
\affil{Department of Particle Physics \& Astrophysics, Weizmann Institute of Science, Rehovot 76100, Israel}

\author{Roger M. Smith}
\affil{Caltech Optical Observatories, California Institute of Technology, Pasadena, CA 91125, USA}

\author{Jamie Soon}
\affil{Research School of Astronomy and Astrophysics, Australian National University, Canberra, ACT 2611, Australia}

\author{Tony Travouillon}
\affil{Research School of Astronomy and Astrophysics, Australian National University, Canberra, ACT 2611, Australia}

\author{Shuangnan Zhang}
\affil{Key Laboratory of Particle Astrophysics, Institute of High Energy Physics, Chinese Academy of Sciences, 19B Yuquan Road, Beijing 100049, China}
\affil{University of Chinese Academy of Sciences, Chinese Academy of Sciences, Beijing 100049, China}

\begin{abstract}
The Galactic magnetar \sgr\ has been reported to produce the first known example of a bright millisecond duration radio burst (FRB\,200428) similar to the cosmological population of fast radio bursts (FRBs), bolstering the association of FRBs to active magnetars. The detection of a coincident bright X-ray burst has revealed the first observed multi-wavelength counterpart of a FRB. However, the search for similar emission at optical wavelengths has been hampered by the high inferred extinction on the line of sight. Here, we present results from the first search for second-timescale emission from the source at near-infrared wavelengths using the Palomar Gattini-IR observing system in $J$-band, made possible by a recently implemented detector read-out mode that allowed for short exposure times of $\approx 0.84$\,s with 99.9\% observing efficiency. With a total observing time of $\approx 12$\,hours ($\approx 47728$\,images) on source during its 2020 outburst, we place median $3\,\sigma$ limits on the second-timescale emission of $\lesssim  20$\,mJy ($13.1$\,AB\,mag). We present non-detection limits from epochs of four simultaneous X-ray bursts detected by the Insight-{\it HXMT} and {\it NuSTAR} telescopes during our observing campaign. The limits translate to an observed fluence limit of $\lesssim 18$\,Jy~ms, while the corresponding extinction corrected limit is $\lesssim 125$\,Jy~ms for an estimated extinction of $A_J = 2.0$\,mag. These limits provide the most stringent constraints (energy $\lesssim 3\times10^{36}$\,erg at 9\,kpc) to date on the fluence of flares at frequencies of $\sim 10^{14}$\,Hz, and constrain the ratio of the near-infrared (NIR) fluence to that of coincident X-ray bursts to $R_{\rm NIR} \lesssim 2.5 \times 10^{-2}$. Our observations were sensitive enough to easily detect a near-infrared counterpart of FRB\,200428 if the NIR emission falls on the same power law as that observed across its radio to X-ray spectrum. The non-detection of NIR emission around the coincident X-ray bursts constrains the fluence index of the brightest burst to be steeper than $\approx 0.35$.
\end{abstract}


\section{Introduction} \label{sec:intro}

The source \sgr\ was discovered in 2014 as a short ($\approx 0.2$\,s) burst \citep{Stamatikos2014} by the Burst Alert Telescope on board the \textit{Neil Gehrels Swift Observatory} \citep{Gehrels2004}. Subsequent follow-up in the X-ray wavebands revealed that the object was a new member of the class of Soft Gamma-ray Repeaters (SGRs) originating from a Galactic magnetar with a spin period of $\approx 3.24$\,s, period derivative of $\dot{P} \approx 1.43 \times 10^{-11}$\,s\,s$^{-1}$, characteristic age of $\approx 3600$\,years and surface magnetic field of $\sim 2 \times 10^{14}$\,G \citep{Israel2016}. The source is coincident with the center of the supernova remnant G57.2+0.8 \citep{Sun2011, Kozlova2016, Zhou2020, Zhong2020} at a distance of $\approx 10$\,kpc. Pulsed radio emission has so far remained undetected at radio bands \citep{Israel2016, Surnis2016, Younes2017, Lin2020b}. In the optical and near-infrared (NIR) regime, a possible faint ($H \approx 24$\,mag) counterpart has been identified in follow-up imaging with the \textit{Hubble Space Telescope} \citep{Levan2018}.

Since its discovery, the source has sporadically gone into outburst over the last few years \citep{Lin2020a}, with the most recent being reported as a ``forest" of X-ray bursts detected during 27-28 April 2020 \citep{Palmer2020, Younes2020}. Following the onset of the outburst, an unprecedented bright millisecond duration radio burst (hereafter FRB\,200428) was detected from the source  by the Canadian Hydrogen Intensity Mapping Experiment (CHIME; \citealt{Chime2020}) and STARE2 \citep{Bochenek2020} telescopes, together with a bright hard X-ray counterpart detected by the {\it INTEGRAL} \citep{Mereghetti2020}, {\it AGILE} \citep{Tavani2020}, {\it Konus-Wind} \citep{Ridnaia2020} and {\it HXMT} \citep{Li2020a} space telescopes. The high luminosity of the radio burst was within a factor of $\approx 40$ of the luminosity of extragalactic Fast Radio Bursts (FRBs; \citealt{Cordes2019, Petroff2019}) observed at cosmological distances, providing strong evidence that at least some FRBs could arise from active SGRs.  

The simultaneous detection of the X-ray burst provides the first evidence of a multi-wavelength counterpart for FRBs. Thus, several optical facilities performed follow-up observations of the source to detect and constrain the presence of optical counterparts coincident with radio/X-ray bursts \citep{Niino2020, Lin2020b}. However, the location of the source in the Galactic plane together with the high observed X-ray column density ($\sim 2 \times 10^{22}$ cm$^{-2}$; \citealt{Israel2016, Younes2017, Li2020a}) suggests a large line-of-sight optical extinction towards the source ($A_V \approx 7 - 10$\,mag). In the case of FRB\,200428, no optical counterpart was detected in a simultaneous observation by the BOOTES telescope \citep{Lin2020b} to an extinction corrected fluence limit of $\lesssim 4400$\,Jy~ms. However, the inferred extinction in the NIR is substantially smaller, and expected to be $\approx 30\%$ of the optical in $J$-band.

Palomar Gattini-IR (PGIR) is a new wide-field NIR time domain survey scanning the entire Northern sky every two nights to a median depth of $J \approx 15.7$\,AB mag \citep{Moore2019, De2020a}. With the implementation of a new detector readout mode that allows for fast (exposure time $\approx 0.84$\,s) and continuous (duty cycle $\approx 100$\%) exposures of the sky, we initiated targeted follow-up observations of the source. In this paper, we describe the PGIR follow-up campaign and constraints from simultaneous NIR observations of \sgr\ at the times of detected X-ray bursts. Section \ref{sec:observations} describes the observing strategy and resulting observation schedule. In Section \ref{sec:data}, we describe the methods used to analyze the acquired data, and Section \ref{sec:results} uses the non-detection of NIR bursts to constrain the fluence ratios of multi-wavelength counterparts of X-ray bursts from \sgr. We conclude with a summary of our results, and prospects for future searches in Section \ref{sec:summary}. 

\section{Observations}
\label{sec:observations}

\begin{table*}[!ht]
    \footnotesize
    \centering
    \begin{tabular}{ccccccccc}
    \hline
    ID & UT Start & UT End & Mode & Exp. time & $N$ & Total exposure & Duty cycle & $3\,\sigma$ limit \\
     & & & & (s) & & (s) & (\%) & (mJy/Jy ms)\\
    \hline
    1 & 2020-05-01 08:24:38.9 & 2020-05-01 12:34:53.6 & I & 1.65 & 2722 & 4491.3 & 29.9 & 9/16\\
    2* & 2020-05-02 07:49:45.5 & 2020-05-02 12:34:14.1 & II & 0.84 & 18009 & 15127.6 & 89.7 & 25/21 \\
    3* & 2020-05-05 08:20:13.7 & 2020-05-05 12:31:09.0 & II & 0.84 & 15917 & 13370.3 & 89.7 &  23/19 \\
    4 & 2020-05-23 11:23:09.0 & 2020-05-23 11:38:27.3 & III & 0.84 & 1084 & 910.6 & 99.9 &  16/13 \\
    5 & 2020-05-24 06:57:05.3 & 2020-05-24 07:29:43.4 & III & 0.84 & 2310 & 1940.4 & 99.9 & 19/16 \\
    6$^{\dagger}$ & 2020-05-24 11:09:27.4 & 2020-05-24 11:34:12.3 & III & 0.84 & 1706 & 1433.0 & 97.3 & 21/18\\
    7 & 2020-05-28 06:13:49.2 & 2020-05-28 07:13:34.7 & III & 0.84 & 4229 & 3552.4 & 99.9 & 18/15 \\
    8$^{\dagger}$ & 2020-05-31 09:01:30.0 & 2020-05-31 09:39:32.9 & III & 0.84 & 1751 & 1470.8 & 64.6 & 62/52\\
    \hline
    \end{tabular}
    \caption{Summary of observing sessions of \sgr\ with PGIR. The Mode column refers the observing configuration of the system during the respective epoch. Mode I indicates the use of the standard observing mode of the survey including dithers between exposures and lower observing efficiency. Mode II indicates the fast readout mode discussed in the text. Both Mode I and II had the telescope aligned to the default observing grid and the source placed away from the best focused part of the field (see discussion of PSF variation in \citealt{De2020a}). Mode III indicates the fast readout mode with the source placed in the best focused part of the detector leading to better sensitivity. $N$ denotes the number of images that produced good quality subtractions in the session. The duty cycle is a conservative lower limit for the first pixel read out in each detector channel. The limiting flux denotes the median limiting flux in the individual exposures at the location of the source as measured from the observed scatter of fluxes in the difference image. Data acquired during epochs marked by * were affected by a bug in the readout that repeated every 10th exposure in the sequence (i.e. every 9th and 10th exposure were identical), leading to reduced duty cycle.  Epochs marked by $^{\dagger}$ were affected by intermittent clouds leading to reduced observing duty cycle. Difference flux measurements and their uncertainties will be released as an electronic supplement upon publication.}
    \label{tab:observations}
\end{table*}

Following the detection of the train of X-ray bursts from \sgr\ \citep{Palmer2020} and FRB\,200428 \citep{Chime2020, Bochenek2020}, we triggered targeted observations of the source using PGIR on UT~2020-05-01. Due to the short expected emission timescale for counterparts from X-ray/radio bursts from the source ($\lesssim 1$\,s) as well as the background noise limited nature of NIR imaging with this instrument (see Table 1 in \citealt{De2020a}), we used the shortest possible exposure time allowed by the standard readout scheme ($1.65$\,s) used in survey operations with an observing efficiency of 30\% including dithers. Following this initial epoch, we significantly increased our observing efficiency, as well as our sensitivity to short timescale flares, by using a newly implemented readout mode of the H2RG detector array in Palomar Gattini-IR \citep{De2020a}. In this new mode, the detector is read out and digitized continuously while exposing on the sky, with an effective exposure time equivalent to the frame readout time of $\approx 0.84$\,s (see Appendix \ref{sec:readout} for details).

Table \ref{tab:observations} provides a summary of all the observing epochs on the source, including the readout mode used and the placement of the source in the large field of view. While the initial observations were designed to monitor the source for the total duration of its night time visibility from Palomar ($\approx 4.5$ hours below airmass of 2) near the peak of its outburst, subsequent epochs in the second half of May 2020 were coordinated with the published visibility windows of the source with the Insight-{\it HXMT} satellite\footnote{Published at \url{http://enghxmt.ihep.ac.cn/dqjh.jhtml}} and the CHIME telescope (K. Smith, pers. comm.).

\section{Data analysis}
\label{sec:data}

\begin{figure}
    \centering
    \includegraphics[width = 0.49\textwidth]{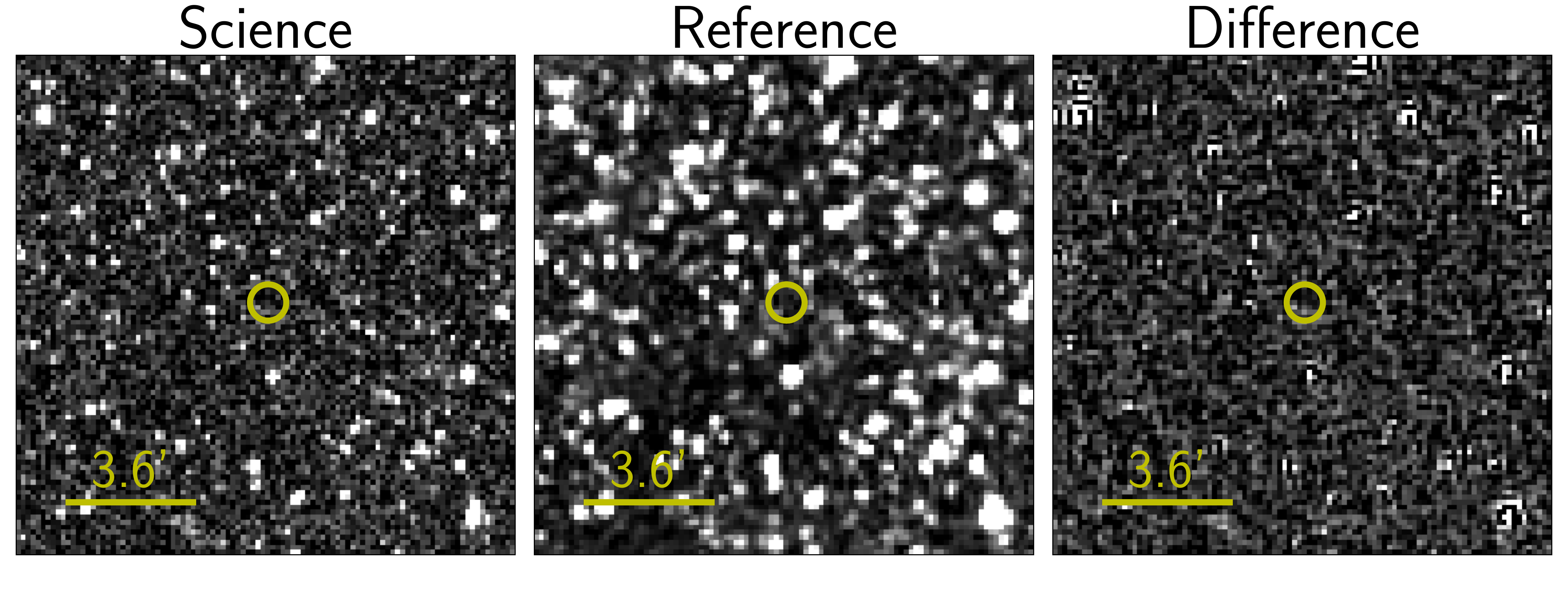}\\
    \includegraphics[width = 0.49\textwidth]{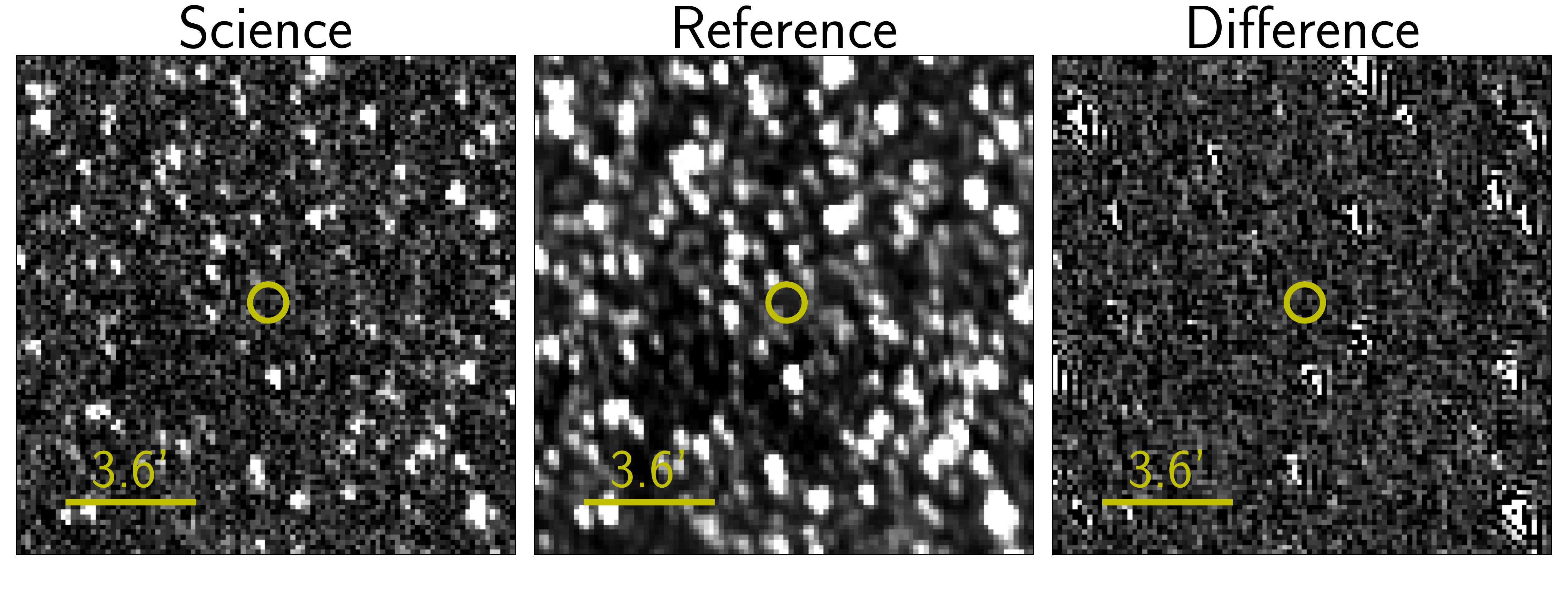}
    \caption{Example cutouts of a science (left column), reference (middle column) and difference image (right column) acquired in our observing sequence. North is up and East is left in each panel. The top and bottom row show examples of the subtractions with the source placed in different parts of the focal plane with differing PSFs -- the top row shows the case where the source was placed in the best part of the detector with approximately symmetric PSFs, while the lower row shows the same in a poorer region of the focal plane with elongated PSFs. In both cases, the difference image produced using ZOGY \citep{Zackay2016} only shows residual astrometric and Poisson noise from nearby bright stars, with no statistically significant transient emission detected at the location of \sgr\ (yellow circle).}
    \label{fig:cutout}
\end{figure}

The location of the source in a dense region of the Galactic plane together with the large pixel scale and under-sampled PSFs of the Gattini observing system present several challenges to the data reduction procedure, which were modified and adapted from the nominal survey mode. Appendix \ref{sec:reduction} provides a detailed summary of the reduction process adopted for this data set.  Figure \ref{fig:cutout} shows an example triplet of a fast readout science frame centered at the location of the source, the corresponding reference image and the resulting difference image. We were able to obtain high quality difference images in all the epochs, which show only background noise fluctuations and residual astrometric/Poisson noise from nearby bright stars.

\begin{figure}
    \centering
    \includegraphics[width=\columnwidth]{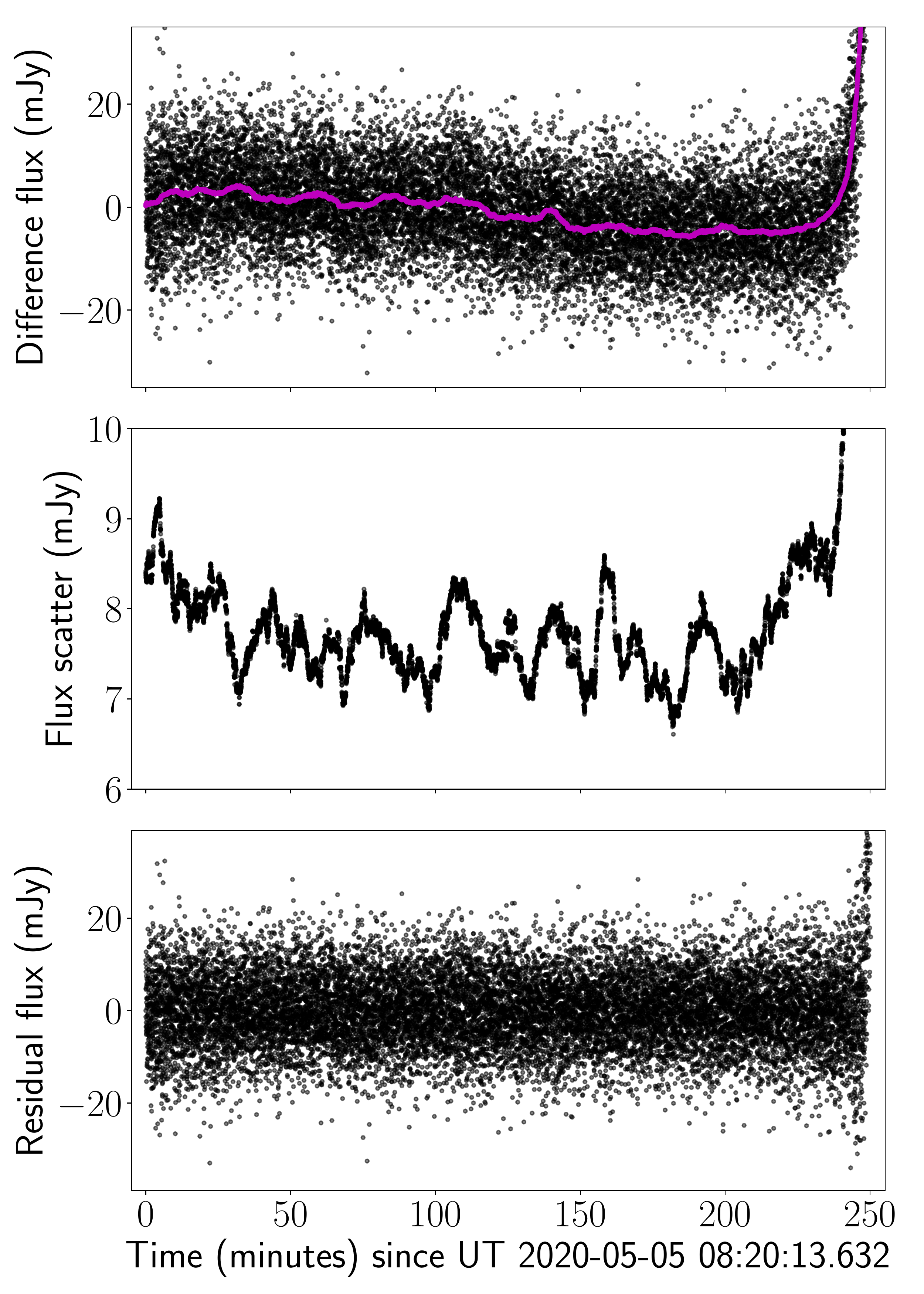}
    \caption{Example of method used to search for second-timescale emission from 
    \sgr\ using Gattini-IR data taken on UT 2020-05-05. (Top panel) Forced photometry time series of fluxes measured at the location of the source in the difference images, with each dot representing a single image and the magneta lines representing a running median of fluxes measured in a window of 200 images in each side. The large increase in the background at the end of the observation is due to the approach of 12 degree twilight at the end of the observation. (Middle panel) Measured standard deviation in fluxes at the location of the source using the same window size as in the running median of the top panel. Noise variations due to the time variable airglow in $J$ band is clearly visible. (Bottom panel) Residual flux obtained from subtracting the longer timescale airglow variations (shown in magneta in the top panel) from the observed flux.}
    \label{fig:fluxseries}
\end{figure}

Figure \ref{fig:fluxseries} shows a time series of the measured difference flux during one of the observing sessions\footnote{The measured flux in counts was converted to mJy using the {\it 2MASS} zero-points published at \url{https://old.ipac.caltech.edu/2mass/releases/allsky/doc/sec6_4a.html}. The corresponding $J=0$ flux density is $1594$\,Jy.}. In order to estimate the uncertainty and signal-to-noise ratio of the flux measurements, we measured the standard deviation of the fluxes in a running window of 200 observations around each image in the sequence. The measured flux scatter exceeds the propagated noise terms by $\approx 10$\% due to the presence of unaccounted noise terms such as correlated noise between the pixels introduced during the resampling process. The measured flux uncertainty exhibits temporal variations of the order of $\approx 20$\% over the duration of the night, reflecting the variation in the foreground $J$-band sky brightness.

In addition to random scatter introduced due to time variable airglow in the $J$-band sky, the measured fluxes also show slow temporal variations (over time scales of tens of minutes) in the median (see Figure \ref{fig:fluxseries}) that correlate with the changing scatter from the sky background, thus arising from imperfect background subtraction with the changing sky background. Since this effect introduces a slow temporal trend, we subtract it using a running median around each image since we aim to detect short timescale flares ($\sim 1$\,s) from the source. The resulting residual time series is shown in Figure \ref{fig:fluxseries}, and is consistent with Gaussian noise in the flux measurements.

\section{Results}
\label{sec:results}

\begin{table*}[]
    \centering
    \footnotesize
    \begin{tabular}{ccccccccc}
    \hline
        ID & Instrument & Trigger time & Duration & X-ray fluence & Obs Start & Obs End & Diff. Fluence & $3\,\sigma$ limit \\
         & & (UT) & (s) & (erg cm$^{-2}$) & (UT Day) & (UT Day) & (erg cm$^{-2}$) & (erg cm$^{-2}$) \\
    \hline
    A & {\it HXMT}/{\it NuSTAR} & 2020-05-02 10:17:26.00 & 0.076 & $7.56 \times 10^{-9}$ & 10:17:25.90 & 10:17:26.74 & $2.69 \times 10^{-12}$ & $4.93 \times 10^{-11}$\\
    B$^\dagger$ & {\it HXMT}/{\it NuSTAR} & 2020-05-02 10:25:25.80 & 0.415 & $1.76 \times 10^{-8}$ & 10:25:25.07 & 10:25:26.77 & $-3.64 \times 10^{-11}$ & $6.49 \times 10^{-11}$ \\
    C & {\it HXMT} & 2020-05-02 10:46:20.85 & 0.077 & $1.16 \times 10^{-10}$ & 10:46:20.12 & 10:46:20.96 & $-7.64 \times 10^{-12}$ & $5.14 \times 10^{-11}$ \\
    D$^\dagger$ & {\it HXMT} & 2020-05-05 12:09:29.75 & 0.039 & $7.97 \times 10^{-9}$ & 12:09:28.94 & 12:09:30.65 & $-1.81 \times 10^{-11}$ & $7.73 \times 10^{-11}$ \\
    \hline
    \end{tabular}
    \caption{List of X-ray bursts reported by high energy instruments within the Gattini-IR observing sequences. The Fluence column denotes the fluence reported by the {\it HXMT} satellite, while the duration denotes their $T_{90}$ measurement. The Obs start and Obs end column denotes the start and end of the exposure that contained the trigger time of the X-ray burst, with respect to the start of the UT day (00:00:00). The Diff. fluence and $3\,\sigma$ limit column denote the IR fluence computed from the difference flux and the corresponding $3\,\sigma$ limit. For bursts denoted by $^\dagger$, the duration of the burst was covered by two consecutive and continuous exposures in the sequence, in which case we reported a weighted flux measurement between the two exposures and its corresponding uncertainty. The IR flux measurements have not yet been corrected for extinction since that is model dependent.}
    \label{tab:bursts}
\end{table*}

Variability in the NIR correlated with X-ray flux changes have been detected in several known Galactic magnetars, but over timescales of days to years (e.g. \citealt{Rea2004, Tam2004, Israel2005}). Fast optical flaring has also been observed in a candidate Galactic soft gamma-ray repeater \citep{Stefanescu2008, Castro-Tirado2008} over timescales of a few seconds. Similarly, a probable faint NIR counterpart (at $H \approx 24$\,mag) of \sgr\ was identified with a deep {\it Hubble Space Telescope} exposure during its 2015 -- 2016 outbursts \citep{Levan2018}, where the IR emission was shown to be enhanced during periods of the X-ray outburst. Yet, it was suggested that the lack of a direct correlation between the NIR -- X-ray fluxes disfavors a disk reprocessing scenario for the NIR emission, and was likely more consistent with a magnetospheric origin of both the emission components \citep{Levan2018}. Here, we focus instead on the detection and limits on second-timescale flares in the NIR, which remain so far observationally unconstrained from this source.

Over the duration of $\approx 12$ hours of observations (Table \ref{tab:observations}), we identified no reliable detections in the NIR time series at a flux level above $3\,\sigma$ from the background noise, beyond that expected from Gaussian noise. The median observed $3\,\sigma$ fluence limit on NIR bursts is $\approx 20$\,Jy~ms (uncorrected for line-of-sight extinction). In order to constrain potential multi-wavelength counterparts, we searched all available public reports of X-ray and radio bursts from the source within our observing time intervals. A total of four X-ray bursts were reported by the {\it HXMT} \citep{CKLi2020}\footnote{The updated list of bursts are available at \url{http://enghxmt.ihep.ac.cn/bfy/331.jhtml}} and NuSTAR \citep{Borghese2020} satellites during our observations. Table \ref{tab:bursts} provides an overview of the X-ray bursts reported during our observations. Notably, no significant emission was found detected around the reported epochs of four X-ray bursts. Below, we use the derived limits from our observations to constrain the fluence ratio of NIR bursts when compared to both the coincident X-ray bursts as well as the observed X-ray to radio spectrum of FRB\,200428.

\subsection{Extinction along the line of sight}

\cite{Israel2016} find the neutral hydrogen column density ($n_{\rm H}$) along the line of sight to be $n_{\rm H} = (1.6 \pm 0.2) \times 10^{22}$\,cm$^{-2}$ using {\it XMM-Newton} spectra fitted by a two-component power-law (PL)
and blackbody (BB) model, from which we obtain an attenuation $A_V = 7.2 \pm 0.9$\,mag \citep{Guver2009} and $A_J = 2.0 \pm 0.3$\,mag \citep{Rieke1985}. This extinction value is consistent with $1.82 < A_J <  1.97$\,mag obtained using 3D dust map based on Pan-STARRS 1, {\it Gaia}, and {\it 2MASS} optical/NIR data \citep{Green2019} assuming a distance of 8--12\,kpc, although these dust maps suffer the lack of bright M-dwarf stars observable at these distances. For $A_J = 2.0 \pm 0.3$\,mag, the corresponding median limits on the intrinsic fluence of the bursts will be $\approx 85 - 150$\,Jy~ms (within a factor of two). For the rest of this work, we assume an extinction of $A_J = 2.0$\,mag towards the source, noting that the exact value does not significantly affect our constraints below due to the smaller effect of extinction in the NIR.

\subsection{Constraints on the NIR fluence ratio from coincident X-ray bursts}

\begin{figure*}[!htp]
    \centering
    \includegraphics[width=0.45\textwidth]{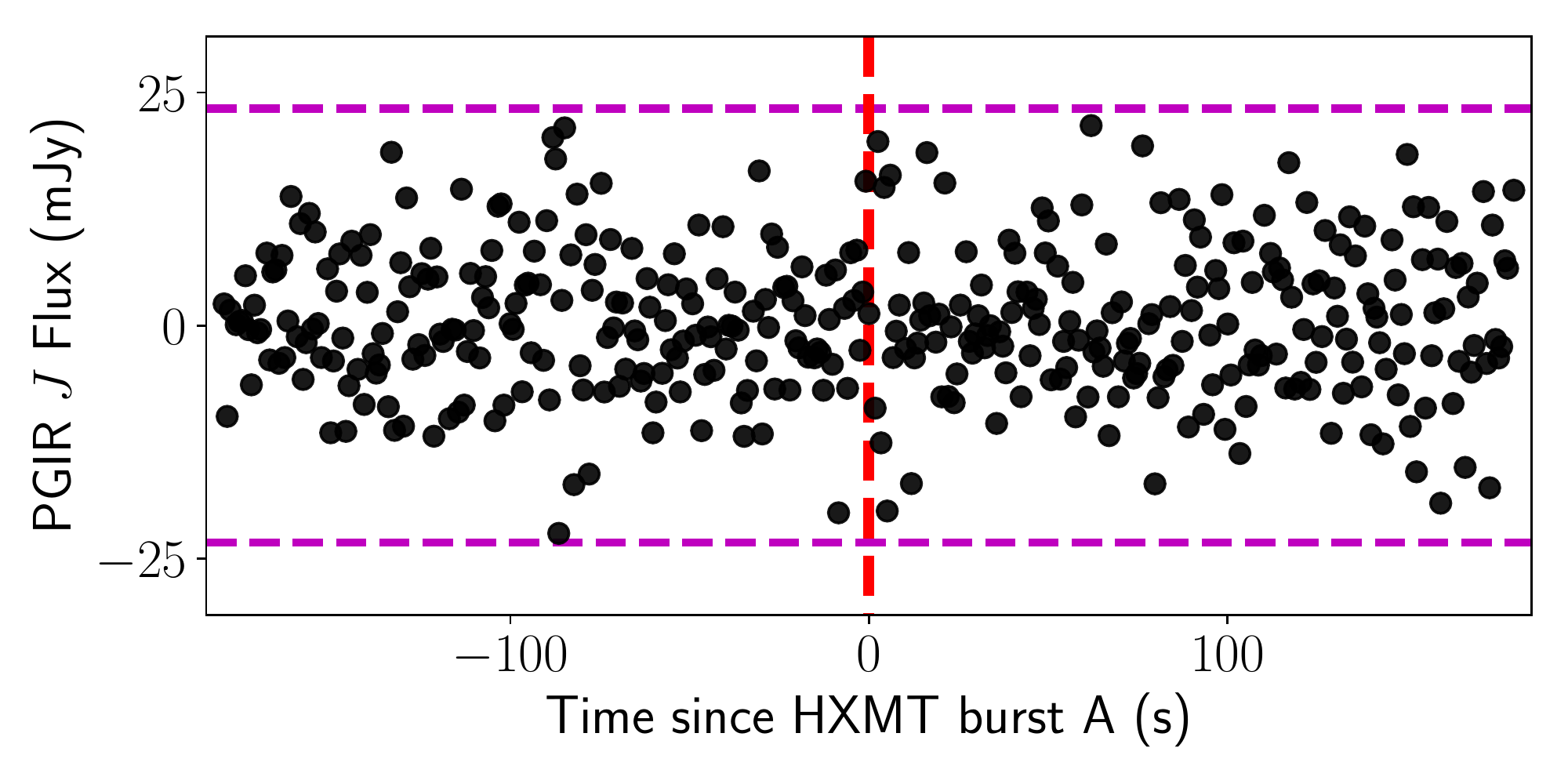}
    \includegraphics[width=0.45\textwidth]{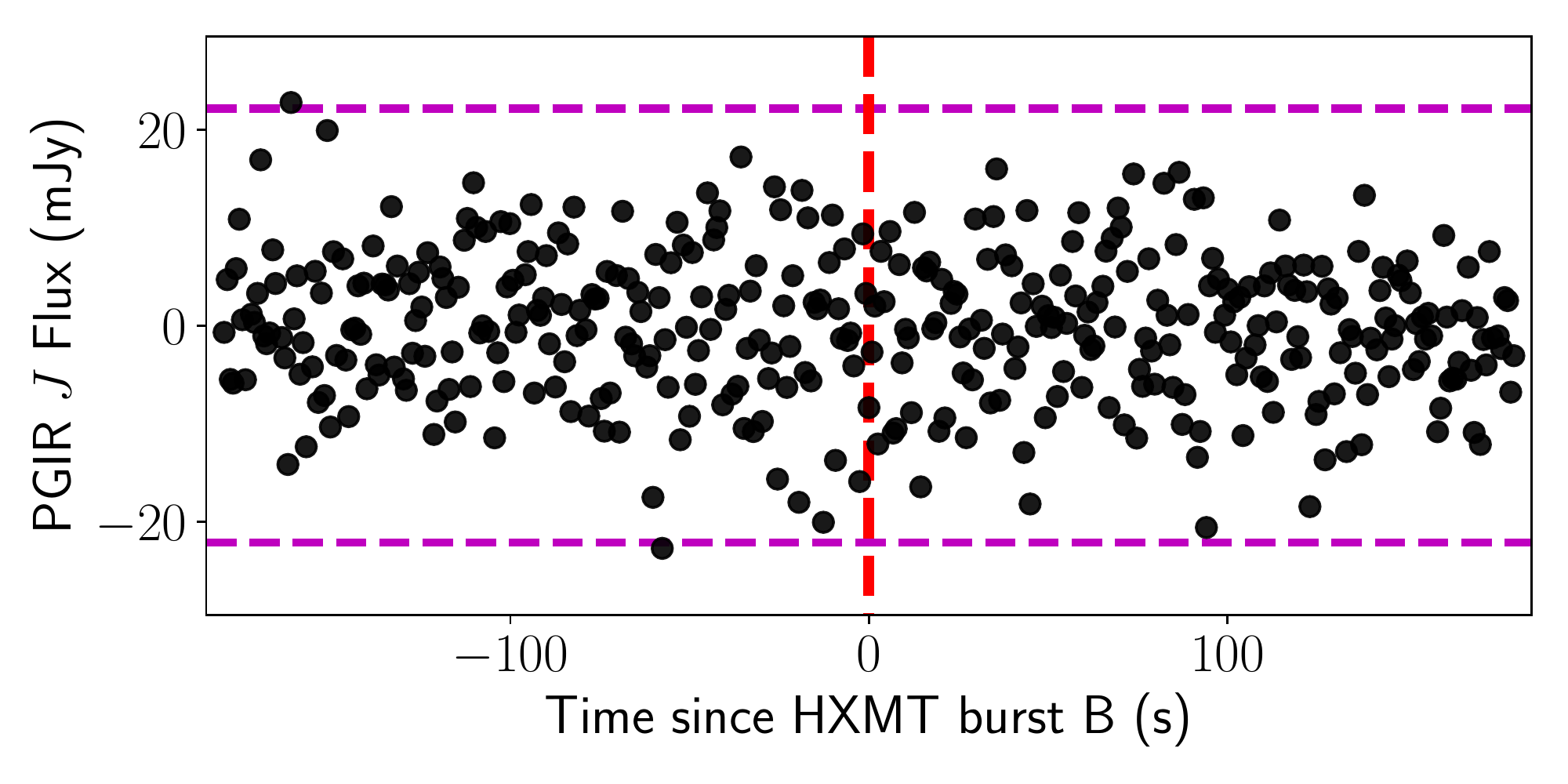}
    \includegraphics[width=0.45\textwidth]{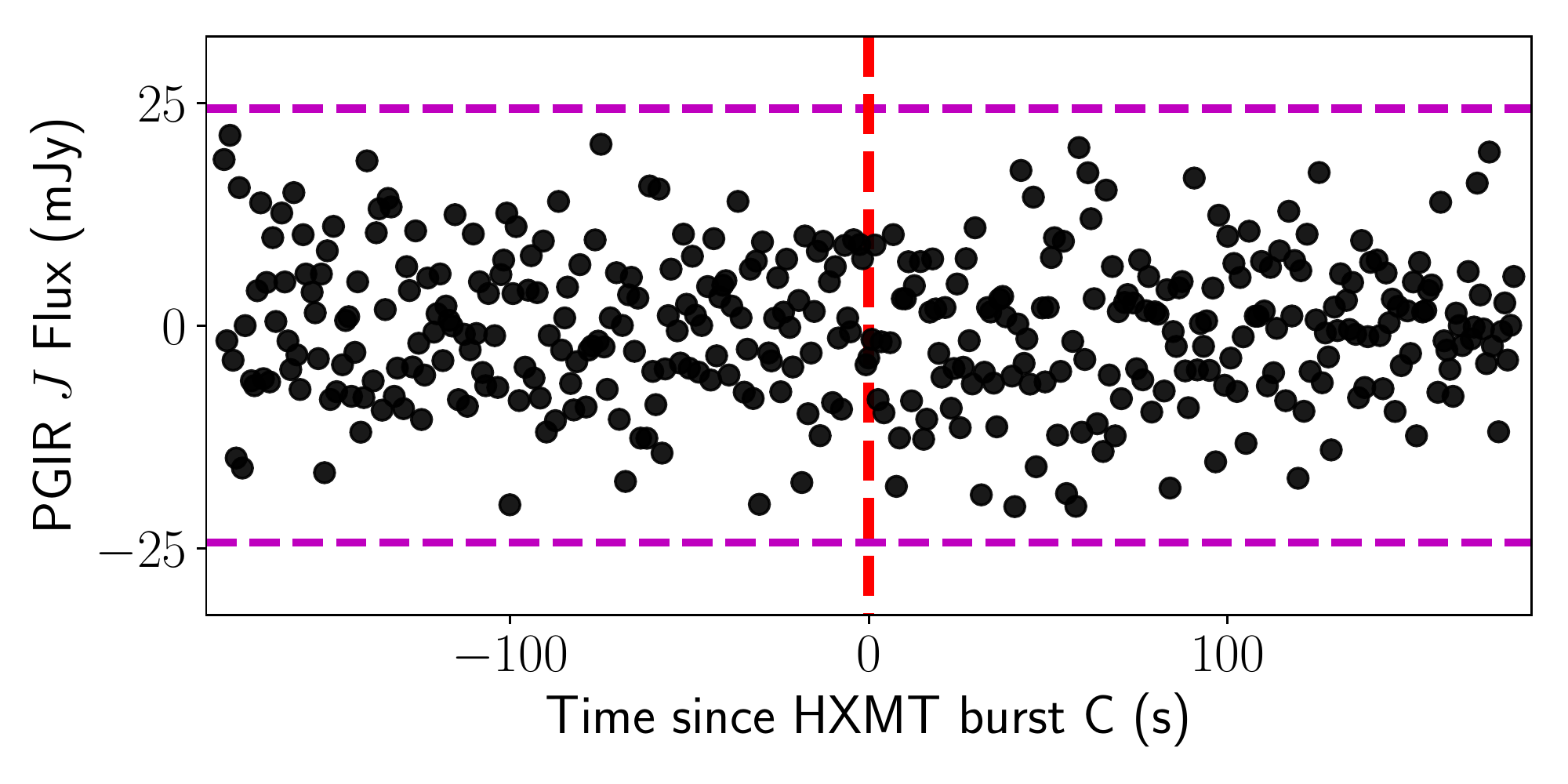}
    \includegraphics[width=0.45\textwidth]{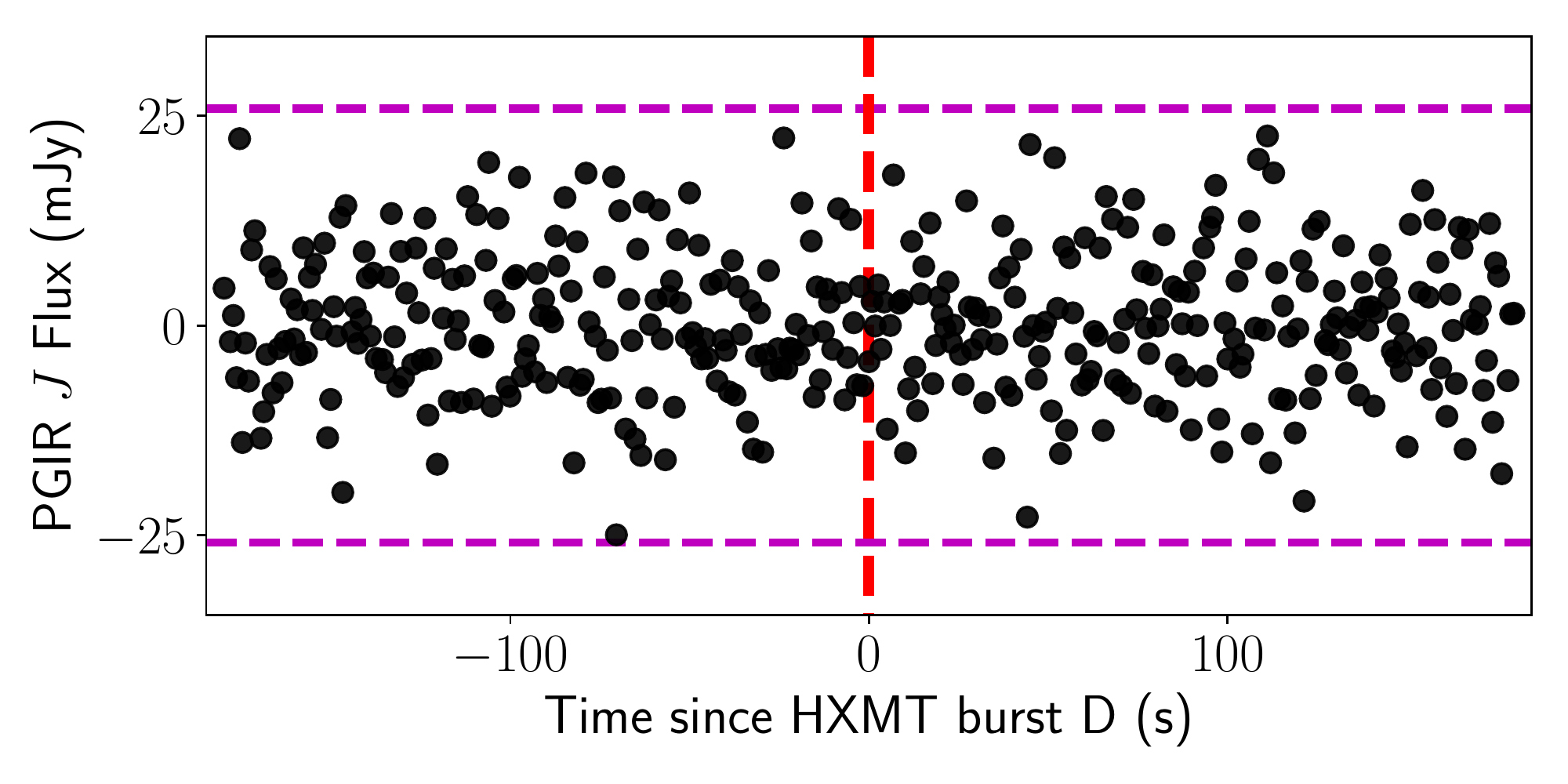}
    \caption{Time series of difference flux measurements over 360\,s intervals centered on the times of known X-ray bursts listed in Table \ref{tab:bursts}. The burst numbering indicated are the same as those in Table \ref{tab:bursts}. The red vertical line shows the time of the reported burst, while the magneta horizontal dashed lines show the $3\,\sigma$ noise levels around the time of observation. No significant emission is detected at the $3\,\sigma$ level around the reported times of the X-ray bursts.}
    \label{fig:bursts}
\end{figure*}

In Table \ref{tab:bursts}, we list the X-ray bursts reported from {\it HXMT} and {\it NuSTAR} during our observing sequence, together with the reported X-ray fluences from {\it HXMT} and our measured difference image flux and corresponding $3\,\sigma$ limit on the NIR fluence\footnote{In order to be consistent with reported X-ray bursts, we define fluence as $\mathcal{F} = \nu f_\nu \delta t$ where $\nu$ is the observed frequency, $f_\nu$ is the spectral flux density and $\delta t$ is the exposure time for our data.}. We note that the expected dispersion delay between X-ray and optical pulses for the reported DM of $\approx 330$ pc cm$^{-3}$ \citep{Chime2020, Bochenek2020, Zhang2020} is $\sim 10^{-11}$\,s and thus not important for our observations. However, since a delay between the X-ray and optical emission could arise as a result of the intrinsic emission mechanism, we show in Figure \ref{fig:bursts}, the measured difference flux in a window of $\approx 6$\,minutes centered on the times of the reported X-ray bursts\footnote{For comparison, we note that the X-ray and radio emission observed in FRB\,200428 was coincident within a maximum conservative uncertainty of $\approx 0.5$\,s, and shorter than our exposure time.}.

No significant emission is identified within this time window of the reported X-ray bursts and we list the derived limits on the NIR fluence of the bursts in Table \ref{tab:bursts}. The strongest constraint on the NIR to X-ray fluence ratio ($R_{\rm NIR}$) is derived from the brightest burst (Burst B), where the non-detection of NIR emission constrains $ R_{\rm NIR} \lesssim 2.5 \times 10^{-2}$ after correcting for extinction. For comparison, we note that the extinction corrected $R_{\rm NIR}$ for longer term correlated X-ray - NIR outbursts (over $\sim$ days -- weeks) observed in Galactic magnetars range from typical values of $\sim 10^{-4}$ (as seen for \sgr\ as well as some other X-ray pulsars; \citealt{Levan2018, Rea2004, Tam2004}) to $\sim 10^{-2}$ (for the IR counterpart of SGR\,1806-20; \citealt{Israel2005}).

\subsection{Comparison to the multi-wavelength properties of FRB\,200428}
\label{sec:frb_comp}

\begin{figure*}[!htp]
    \centering
    \includegraphics[width=0.75\textwidth]{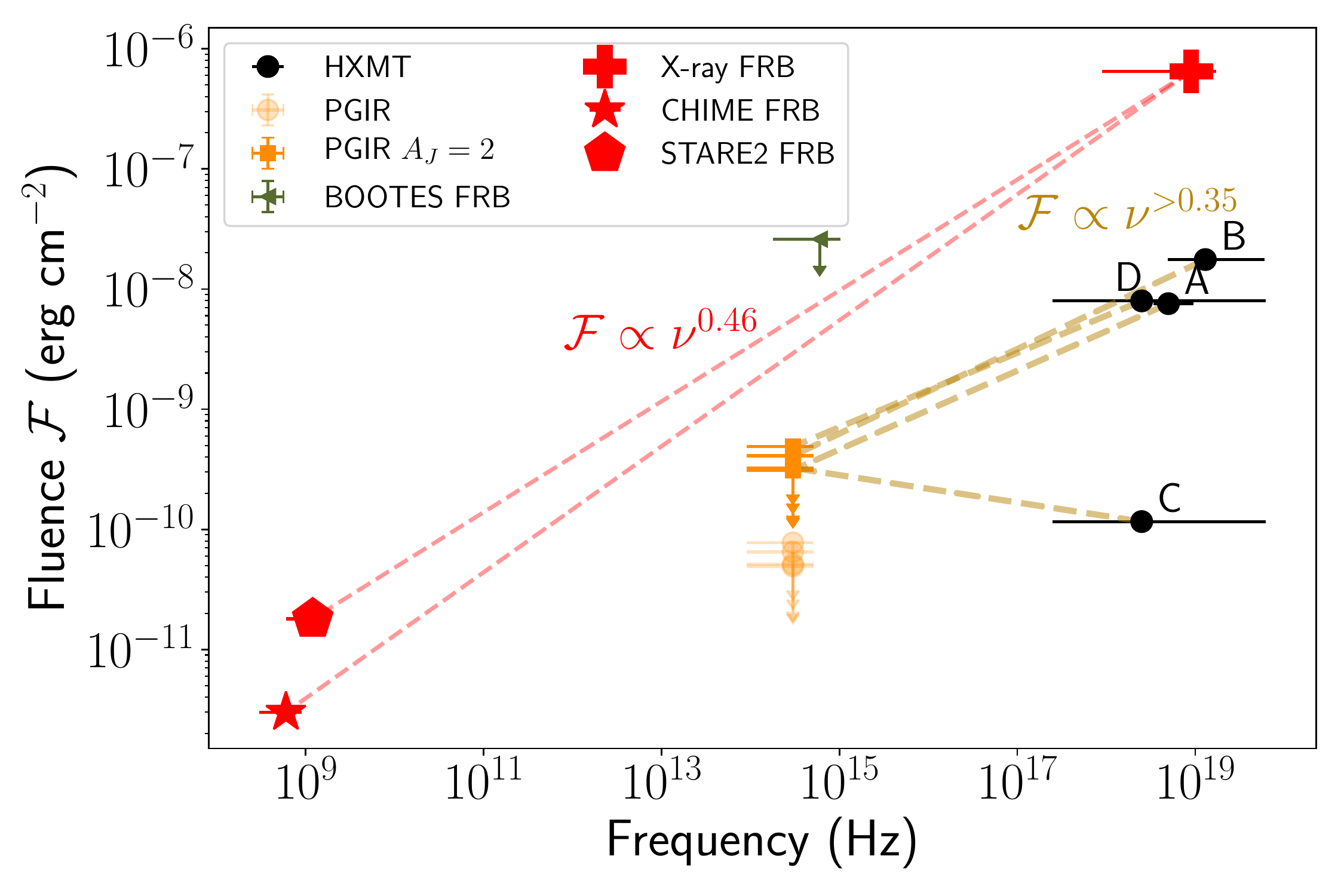}
    \caption{Constraints on the NIR fluence of X-ray bursts from \sgr\ based on limits from Palomar Gattini-IR. For comparison, we show the observed X-ray and radio fluence of FRB\,200428 (in red) detected by CHIME and STARE2 in coincidence with a hard X-ray burst detected by {\it HXMT}, {\it INTEGRAL}, {\it AGILE} and {\it Konus-Wind}. Optical limit from the BOOTES telescope for FRB\,200428 is also shown. Limits from the PGIR campaign are shown in orange, with transparent circles showing raw limits, while solid squares show extinction corrected limits for $A_J = 2.0$\,mag. The four X-ray bursts detected by {\it HXMT} and {\it NuSTAR} during the PGIR observing sessions (labels indicated as per Table \ref{tab:bursts}) are shown with yellow dashed lines connecting the corresponding NIR limits. For each X-ray burst, we place the fluence at the frequency corresponding to the peak of the fluence  spectrum ($= \nu f_\nu$) observed in the X-rays (and its uncertainty). The estimated fluence peak for bursts A and B are consistent with the observed spectrum of the bursts from NuSTAR observations \citep{Borghese2020}, where the fluence is observed to rise up to at least $\sim 20$\,keV. In the case of burst C and D, the fluence spectrum was not well constrained and hence are placed nominally at 10 keV with error-bars spanning the full HXMT sensitivity range. The observed fluence index for FRB\,200428, and constraints derived for the X-ray bursts are shown.}
    \label{fig:sed}
\end{figure*}

As the only other X-ray burst from \sgr\ that has been reported with a multi-wavelength (radio) counterpart, we compare the NIR limits to the observed spectral energy distribution (SED) of FRB\,200428. The striking time coincidence between two X-ray pulses observed in the X-ray burst associated with FRB\,200428 \citep{Li2020a, Mereghetti2020} with the two radio pulses detected by CHIME \citep{Chime2020} potentially suggest a common emission source extending from the X-ray to radio frequencies. \citet{Li2020a} show that the HXMT X-ray burst associated FRB\,200428 was characterized with a hard power law spectrum with a photon index\footnote{The corresponding flux density dependence is $f_\nu \propto \nu^{-0.5}$ and fluence dependence is $\mathcal{F} \propto \nu^{0.5}$ } of $\Gamma \approx 1.5$. In particular, they show that the observed STARE2 fluence at $\sim 1.4$\,GHz can be explained by a single power law in $f_\nu$ extending from X-ray to radio frequencies (see also \citealt{Ridnaia2020}).

In Figure \ref{fig:sed}, we show the observed fluence of FRB\,200428 as a function of frequency, which suggests a fluence dependence of approximately $\mathcal{F} \propto \nu^{0.46}$. In addition, Figure \ref{fig:sed} shows the observed fluences of the X-ray bursts reported within our observing session along with our NIR fluence limits. While the spectrum of the X-ray burst for FRB\,200428 remains unconstrained below $\sim 1$\,keV, Figure \ref{fig:sed} shows that our observations had the sensitivity to detect (at $\sim 30\,\sigma$ confidence) a NIR counterpart of FRB\,200428 if PGIR was observing at the time of the burst and the emission was characterized by a continuous power law extending from X-ray to radio frequencies. We note that the BOOTES limit from \citealt{Lin2020b} does not rule this out. However, we caution that the observed radio spectrum of FRB\,200428 shows signatures of narrow bandwidth fluctuations \citep{Chime2020, Bochenek2020}. Hence, the SED of FRB\,200428 may not be consistent with a single power law extending from X-ray to radio frequencies, although propagation effects may affect this interpretation.

We compare this fluence index\footnote{Here we refer to the fluence index as the exponential factor $\beta$ of the observed fluence that scales as $\mathcal{F}\propto\nu^\beta$.} ($\beta$) to the expected NIR emission from the coincident X-ray bursts within our observing sessions. The strongest constraints on the fluence index are derived from the brightest burst (Burst B), for which the fluence index\footnote{The corresponding constraint on the spectral index $\alpha$ of $f_\nu$ is $\alpha > -0.65$} is constrained to $\beta > 0.35$. This index is close but not constraining enough to rule out a NIR counterpart for these X-ray bursts with the estimated fluence index of FRB\,200428, assuming the NIR emission falls on the same power law as observed across the X-ray to radio spectrum. The non-detection is consistent with radio constraints derived from the non-detection of radio bursts by FAST of $29$ bursts from \sgr\ detected by Fermi-GBM \citep{Lin2020b}, who derive deep limits of $\sim 0.03$\,Jy~ms at 1.25~GHz for these bursts. These non-detections require much steeper X-ray to radio fluence indices ($\beta > 1.2$) for the majority of bursts from SGR\,1935+2154, suggesting that our limits in the NIR would not be deep enough to detect possible counterparts of the majority of bursts.

\subsection{Comparison to theoretical models}
Recent works have aimed to provide constraints on several proposed models for FRBs to explain the observed occurrence of FRB\,200428 simultaneously with the bright X-ray burst (e.g. \citealt{Margalit2020, Lu2020}). These models primarily revolve around two scenarios -- one where the X-ray/radio emission is generated inside the neutron star magnetosphere via coherent curvature radiation (e.g. \citealt{Pen2015, Cordes2016, Kumar2017, Lu2018}) or via coherent maser processes produced at shock interaction of relativistic ejecta  with circumstellar material (e.g. \citealt{Lyubarsky2014, Beloborodov2019, Metzger2019, Margalit2020}). In addition, the X-ray, optical/NIR and radio emission may not be generated at the same location near the neutron star in several of these scenarios. Since the theoretical predictions for multi-wavelength counterparts are not well constrained enough to interpret our upper limits, we only briefly compare them to our NIR observational data.

\citet{Chen2020} provide a summary of the predictions for the fluence in the aforementioned models. In the case of the relativistic shock interaction model by \citet{Beloborodov2019}, if the the blast wave strikes a wind bubble in the tail of a previous flare, a bright optical flare could result with an optical to radio fluence ratio of $\mathcal{F}_{\rm opt} / \mathcal{F}_{\rm radio} \lesssim 10^5$ \citep{Chen2020}. If some X-ray bursts from \sgr\ during the PGIR campaign were accompanied by a radio burst similar to FRB\,200428, then we have the corresponding prediction of $R_{\rm NIR} \lesssim 0.1$ for $\mathcal{F}_{\rm radio} / \mathcal{F}_{\rm X-ray} \sim 10^{-6}$ (as observed by STARE2; \citealt{Bochenek2020}) and $R_{\rm NIR} \lesssim 0.01$ for $\mathcal{F}_{\rm radio} / \mathcal{F}_{\rm X-ray} \sim 10^{-7}$ (as observed by CHIME; \citealt{Chime2020}). Our upper limits are thus comparable to these model predictions for the brighter X-ray bursts. On the other hand, for the curvature radiation model of \citet{Lu2018}, very little NIR/optical (or higher-frequency) emission is expected from the coherently emitting particles. The transient event may
be accompanied by incoherent emission processes inside the magnetosphere, and the maximum possible NIR flux from any incoherent emission processes from an emitting volume of radius $r=10^8 r_8$\,cm and plasma temperature $T=10^8 T_8$\,K is given by
\begin{eqnarray}
    \mathcal{F}_{max} \sim (h\nu/kT)^3 \sigma T^4 (r/D)^2 \\
    \sim 10^{-13}
 T_8 r_8^2\, {\rm erg\,cm^{-2}\,s^{-1}}
\end{eqnarray}
where $D \sim 9$\,kpc is the distance to the source. We see that the
NIR/optical emission from within the magnetosphere is undetectable in our
observations for burst duration $\lesssim 1$\,s. 

\section{Summary}
\label{sec:summary}

In this Letter, we have presented results from a targeted follow-up campaign to search for second timescale NIR flares from the Galactic magnetar \sgr\ using Palomar Gattini-IR. The observations were enabled with a recently implemented detector read-out mode that allows for high time resolution readout of the detector array with nearly 100\% observing efficiency. We found no significant counterparts for second timescale flares from the source above a median $3\,\sigma$ fluence limit of $\approx 20$\,Jy ms. This non-detection, together with the relatively low inferred extinction towards the source at NIR wavelengths ($A_J \approx 2.0 \pm 0.3$\,mag) allows us to place the most stringent extinction-corrected constraints till date on second-timescale flares from the source of $\approx 85 - 150$\,Jy~ms at optical/NIR wavelengths ($\nu \sim 10^{14}$\,Hz). The NIR limit corresponds to an energy $\rm{E} \lesssim 3\times 10^{36}$\,erg at a distance of 9\,kpc \citep{Zhong2020}, and is within an order of magnitude of that reported in the radio for FRB\,200428 at 1.25\,GHz ($\approx 2 \times 10^{35}$ erg; \citealt{Bochenek2020}). It is also several orders of magnitude deeper than reported optical limits from nearby well-localized FRBs ($\sim 10^{43 - 46}$\,erg; \citealt{Hardy2017, Andreoni2020}).

A total of four X-ray bursts were detected by the {\it HXMT} and {\it NuSTAR} telescopes within our continuous observing campaign, although no NIR counterparts were detected. The non-detection of NIR emission around these bursts constrain the NIR to X-ray fluence ratio to $R_{\rm NIR} \lesssim 2.5 \times 10^{-2}$. Comparing these fluence limits to the radio/X-ray fluence observed in FRB\,200428, we show that our observations were sensitive enough to detect a NIR counterpart at a significance of $\sim 30\,\sigma$ if PGIR was observing at the time of FRB\,200428 and the NIR emission falls on the same power law suggested for the radio/X-ray emission. The non-detection of NIR emission associated with the brightest X-ray burst within our observation time constrains the X-ray to NIR fluence index of the burst to be $\beta > 0.35$ (spectral index $\alpha > -0.65$).

As Palomar Gattini-IR performs the first all-sky untargeted time domain survey of the dynamic infrared sky at timescales of days to years over the survey duration, these observations further demonstrate a unique new capability of this instrument to probe the infrared time domain sky at second timescales. Although the instrument uses a small (30\,cm) telescope with coarse pixels that severely limit its sensitivity due to the bright $J$-band foreground, these observations prove the scientific utility of specialized NIR detector read-out modes in finding large amplitude second-timescale flares from dust-obscured sources in the Galactic plane. This advocates for a systematic exploration of this hitherto unexplored phase space, which is possible with PGIR not only for single sources (as demonstrated in this work) but for large patches of the sky, enabled by the instrument's large field of view. 

\acknowledgements
We thank the Insight/HXMT team for their kind co-operation in co-ordinating observations and quickly providing fluence estimates. We thank K. Smith for cooperation regarding the CHIME observability windows. We thank C. Bochenek, W. Lu, V. Ravi and S. R. Kulkarni for valuable discussions on this work.

Palomar Gattini-IR (PGIR) is generously funded by Caltech, Australian National University, the Mt Cuba Foundation, the Heising Simons Foundation, the Bi-national Science Foundation. PGIR is a collaborative project among Caltech, Australian National University, University of New South Wales, Columbia University and the Weizmann Institute of Science. This work was supported by the GROWTH (Global Relay of Observatories Watching Transients Happen) project funded by the National Science Foundation under PIRE Grant No 1545949. GROWTH is a collaborative project among the California Institute of Technology (USA), University of Maryland College Park (USA), University of Wisconsin Milwaukee (USA), Texas Tech University (USA), San Diego State University (USA), University of Washington (USA), Los Alamos National Laboratory (USA), Tokyo Institute of Technology (Japan), National Central University (Taiwan), Indian Institute of Astrophysics (India), Indian Institute of Technology Bombay (India), Weizmann Institute of Science (Israel), The Oskar Klein Centre at Stockholm University (Sweden), Humboldt University (Germany), Liverpool John Moores University (UK) and University of Sydney (Australia).  

\bibliography{sample63}

\appendix

\section{Fast readout mode}
\label{sec:readout}
Conventional readout in the H2RG array requires one frame scan to reset and measure the resulting initial offset (which contains various random errors) and a second frame to measure the final value.  The signal is then the difference of these two frames.  At the fastest frame rate (no delay between frames), the duty cycle drops to 50\%. In the new fast readout mode, we altered the readout sequence such that each line is digitized, then reset and digitized again before preceding to the next line. Signal is then being recorded except during the interval between signal and post-reset level samples. Given the high sky noise, we were able to reduce signal sampling time so that pixel time was reduced from typical $\approx 6 - 7\,\mu$s to $3.1\,\mu$s.  With two samples per pixel per frame, the frame time is then $0.848039$\,s.  The dead time between reading final sample and the next  post-reset sample $\approx 200\,\mu$s. The first frame in a sequence still requires two frames, with the first merely establishing the post reset level. The duty cycle for an $N$ frame sequence is then
\begin{equation}
   	\frac{N-1}{N} \times \frac{0.848039 - 0.0002}{0.848039}
\end{equation}
which approaches $99.98$\% for long exposure sequences. Anomalous behavior due to self heating variations were avoided by clocking the detector continuously. The camera was setup to read continuously and store data in a ten frame circular buffer in the computer's memory with the only distinction between “idling” and exposing being whether the data was written to disk.

\section{Data reduction}
\label{sec:reduction}

While the nominal survey mode operations in Gattini-IR use the \texttt{Drizzle} \citep{Fruchter2002} technique to reconstruct the under-sampled PSFs by stacking several dithered images taken on sky, our requirement for high time resolution at the native image readout timescale makes it unsuitable for this application. We thus modified our default data processing pipeline to perform detrending, astrometry, photometry and subtractions on individual images at the native pixel scale of the detector, which we describe below.

\subsection{Flat-field generation and image detrending}

A master flat-field was created for the read-out mode using a median combination of $400$ sky images across several observing nights in order to calibrate the pixel-to-pixel response of the array in the new readout mode. Using images acquired a wide range of times ensures that temporal structures in the sky background variation do not affect the resulting flat-field. Each acquired image ($2048\times 2048$ pixels) was flat-fielded using the derived flat-field, and only 1/16 of the full image ($512 \times 512$ pixels; hereafter referred to as a sub-quadrant as per the terminology in \citealt{De2020a}) containing the target of interest was retained for further processing. Retaining a smaller portion of the image containing the source leads to a large reduction in the variation of the PSF across the image, thus producing better quality astrometric and photometric solutions, as well as subtractions downstream.

\subsection{Astrometry, photometry and reference image generation}

An astrometric and photometric solution for the sub-quadrant was derived using relatively bright and isolated stars in the field. The calibration was performed using the same techniques as in the regular Gattini observing system and using the same reference catalog, which is calibrated astrometrically to {\it Gaia} DR2 and photometrically to the {\it 2MASS} point source catalog. The astrometric solutions achieved typical RMS of $\approx 1.0 - 1.2$\,\arcsec ($\lesssim 1/8$ of a pixel) , while the photometric solutions have typical uncertainties on the zero-point of $\approx 1$\% as calibrated from $\sim 100$ stars in each image. 

Due to the location of the source in a dense region of the Galactic plane (see Figure \ref{fig:cutout}) and the highly non-stationary background limited by confusion noise, direct aperture photometry measurements on un-subtracted images are not well-suited for deriving accurate constraints on the source flux. We thus created for each observing night, a deep median stack of $400$ sub-quadrants to serve as a reference image with nearly identical PSF as the science images taken each night. Since the reference stack was produced as a median combination as implemented in \texttt{Swarp} (\citealt{Bertin2002}; at the same pixel scale as that of the science images), we do not expect any short timescale emission to contaminate the reference image.

\subsection{Difference imaging and forced photometry}

Each reduced sub-quadrant was processed through image subtraction by resampling the respective reference image (one for each night) to the coordinate grid of the science frame. The resampled reference frame was then flux-scaling to each science frame using common cross-matched stars in the two images. The typical astrometric registration uncertainty between the cross-matched stars was $\approx 0.1 - 0.15$\,pixels, while the corresponding flux scaling certainty was $\lesssim 5$\%. Image subtraction was performed using the ZOGY algorithm \citep{Zackay2016}, including propagation of noise uncertainties from the science and reference images to produce an uncertainty image for each produced difference image (as in \citealt{De2020b}). The flux and its uncertainty at the location of the source was measured directly from the difference images by performing a weighted flux measurement using the difference image PSF at the location of the target in the difference image and the corresponding uncertainty image.

\end{document}